\documentclass[letterpaper, 10 pt, conference]{ieeeconf}  % Comment this line out if you need a4paper

\IEEEoverridecommandlockouts                              % This command is only needed if 
\usepackage{cite}
\usepackage{amsmath,amssymb,amsfonts}
\usepackage{algorithmic}
\usepackage{graphicx}
\usepackage{textcomp}
\usepackage{xcolor}
\usepackage[caption=false,font=footnotesize]{subfig}
\usepackage{booktabs}
\usepackage{multirow}
\usepackage{makecell}
\usepackage{amsmath}

\title{\LARGE \bf
Gender Fairness in Audio Deepfake Detection: Performance and Disparity Analysis
}

\author{
    Aishwarya Fursule$^{1}$,
    Shruti Kshirsagar$^{1}$,
    and Anderson R. Avila$^{2,3}$%
    \thanks{$^{1}$School of Computing, Wichita State University, Wichita, KS, USA}%
    \thanks{$^{2}$Institut national de la recherche scientifique (INRS--EMT), Montreal, QC, Canada}%
    \thanks{$^{3}$INRS-UQO Mixed Research Unit on Cybersecurity, Gatineau, Canada}%
    \thanks{Emails: \{axfursule@shockers.wichita.edu,shruti.kshirsagar@wichita.edu, anderson.avila@inrs.ca\}}
}

\begin{document}
\maketitle
\thispagestyle{empty} 
\pagestyle{empty}

%%%%%%%%%%%%%%%%%%%%%%%%%%%%%%%%%%%%%%%%%%%%%%%%%%%%%%%%%%%%%%%%%%%%%%%%%%%%%%%%
\begin{abstract}

Audio deepfake detection aims to distinguish real human voices from those generated by Artificial Intelligence (AI) and has become an important problem for voice biometrics systems. As synthetic speech continues to improve in quality and realism, its potential misuse for malicious activities, such as identity theft, poses growing security risks. Despite recent progress in the field, the issue of gender bias remains underexplored. In this paper, we present a systematic analysis of gender fairness in audio deepfake detection. Alternatively to conventional evaluation practices based on Equal Error Rate (EER), we adopt five group fairness metrics to quantify gender disparities across different models. We conduct experiments on the ASVspoof 5 dataset using a ResNet-18 classifier with four audio feature representations, and compare them against the AASIST baseline. Our results highlight statistically significant gender disparities across multiple fairness criteria, revealing differences in error distribution that are not characterized by aggregate metrics such as EER. These findings demonstrate that relying solely on standard performance measures is not sufficient to assess model fairness, underscoring the importance of fairness-aware evaluation for developing more equitable, robust, and trustworthy audio deepfake detection systems.

\end{abstract}

%%%%%%%%%%%%%%%%%%%%%%%%%%%%%%%%%%%%%%%%%%%%%%%%%%%%%%%%%%%%%%%%%%%%%%%%%%%%%%%%
\section{Introduction}

Audio deepfakes have become more sophisticated due to recent advances in Artificial Intelligence (AI) and deep learning techniques \cite{c1}. Generative models can now create synthetic speech that sounds natural and resembles the human voice, making it difficult to distinguish between real and AI-generated content \cite{c2}. As a result, audio deepfake models have become valuable tools for bad actors, who may use them for malicious purposes, such as identity theft and the spread of spoken misinformation \cite{c3}. As a countermeasure, several audio deepfake detection approaches have been proposed in the literature \cite{c4, c13, c120, c121, c122}, with most of them evaluated on benchmark datasets, such as the Automatic Speaker Verification and Spoofing Countermeasures Challenge (ASVSpoof) \cite{c108}. While these initiatives offer a reliable testbed for comparative analysis between new detection solutions, they often lack a systematic examination of how performance differs \cite{c13, c12, c106}. 

%In fact, most studies about fairness in deepfake detection are focused on image and video modalities, with limited work based on audio. 

To date, fairness assessment has primarily focused on image and video modalities, and investigations in audio deepfake detection systems, particularly for gender bias, have only recently emerged. The authors in \cite{c106}, for instance, evaluated machine learning and deep learning models on a gender-balanced audio deepfake dataset and reported differences in performance for male and female speech. The authors report higher detection accuracy for female voices across several configurations, providing empirical evidence that gender characteristics influence audio deepfake detection performance. In the study presented in \cite{yadav2024fairssd}, a framework (namely FairSSD) is proposed to evaluate bias in an audio deepfake detection system. Their analysis revealed that existing detectors exhibit significant gender bias, with systematically higher false positive rates for male speakers compared to the female ones. In another study \cite{bird2023real}, it was reported that the models trained on female audio-only outperformed those trained on male audio-only, suggesting that the expressiveness of female voices and the presence of high-pitched artifacts in synthetic speech lead to better detectability among female voices. In a recent study \cite {yang2025}, the authors systematically assessed demographic bias across gender, age, accent, and language, showing that detection accuracy varies significantly by speaker gender. These biases are amplified by training on imbalanced corpora, leading to unequal error rates across speaker groups. 

It is known that speech signals naturally differ between male and female speakers in pitch, vocal range, and speaking patterns \cite{c8}. Thus, a model trained under the same regime for different voice types is expected to yield group-dependent performance. If such variations are not considered during training, the detection model may become biased, exhibiting uneven performance across genders \cite{c9, panda2026bias}. This underscores the importance of systematic fairness evaluation \cite{c11,c124}, assuring that deployed systems operate equitably across all user groups \cite{c116, c7}. Nevertheless, existing audio deepfake detection studies rarely incorporate fairness metrics in their evaluation, relying on traditional performance measures, such as accuracy or EER, which do not explicitly quantify how errors are distributed across demographic groups. Therefore, fairness metrics are needed to capture group-dependent disparities that may remain hidden under traditional measures. This study aims to fill this gap by investigating gender fairness in audio deepfake detection using five fairness metrics, i.e., Statistical Parity, Equal Opportunity, Equality of Odds, Predictive Parity, and Treatment Equality \cite{c16}, which were essential to capture uneven performance beyond traditional measures, such as the Equal Error Rate (EER). We adopt the most recent ASVSpoof Challenge dataset and evaluate four feature representations based on the ResNet-18 classifier \cite{c113}. Additionally, a state-of-the-art model, namely AASIST \cite{AASIST2025}, was also assessed as an end-to-end solution. Our experiments show that performance is influenced by gender for scenarios where the same experimental settings are considered.

\section{Methodology}
\label{method}
%This section describes the fairness metrics used to evaluate gender-based disparities in audio deepfake detection. Standard fairness metrics provide quantitative measures of model behavior across demographic groups and indicate whether the outcomes are equitable, although they do not directly explain the underlying causes of unfairness. These metrics offer critical insight into performance disparities and motivate further analysis and improvement. Section~\ref{fairness} introduces the standard fairness metrics used in this work, while Section~\ref{fairness1} explains how these metrics are instantiated and applied to gender-aware audio deepfake detection.

\begin{figure}[t]
    \centering
    \includegraphics[width=\linewidth]{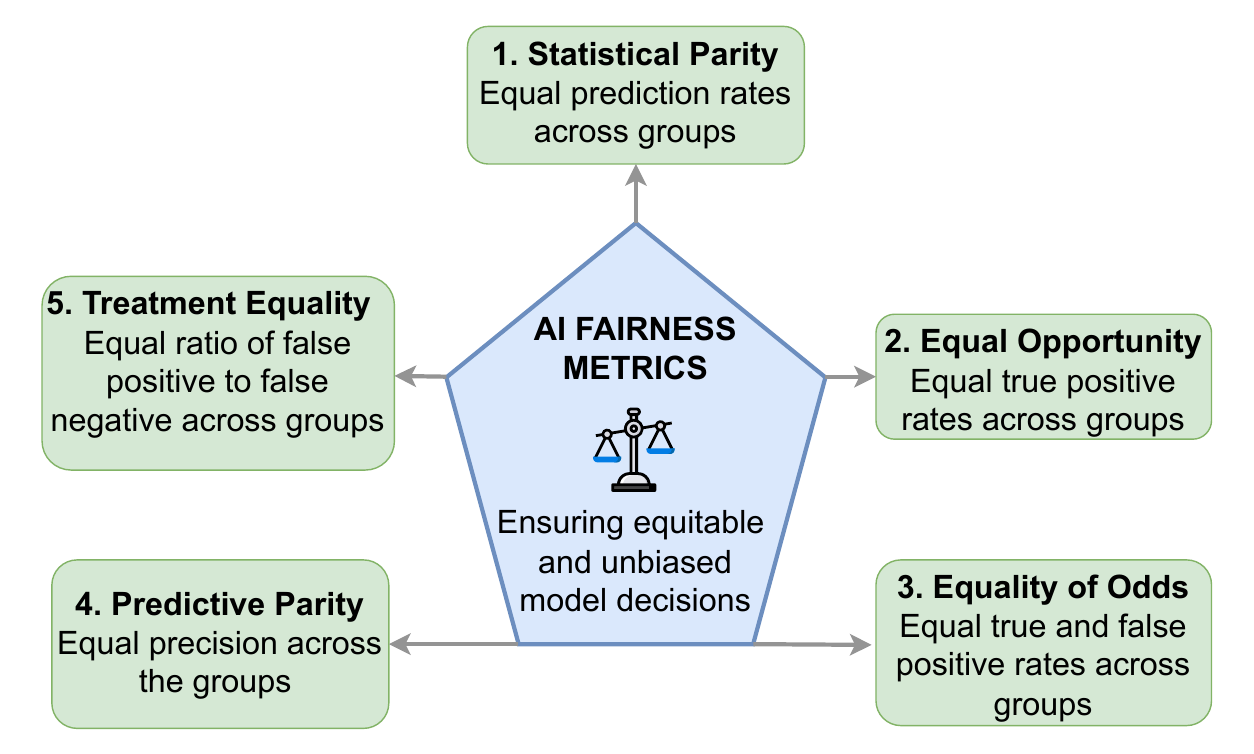}
    \caption{The five group fairness metrics adopted in this study.}
    \label{fig:fairnessmetrics}
\end{figure}

In this study, we evaluate audio deepfake fairness with respect to two demographic groups (i.e., female and male speakers). The task is formulated as a binary classification problem, with $Y=1$ denoting AI-generated speech, and $Y=0$ representing bonafide speech. The predicted class is denoted by $\hat{Y}$. Five fairness metrics (see Figure~\ref{fig:fairnessmetrics}) are adopted to quantify model behavior across gender groups (represented here by $G$), where $g \in \{f, m\}$ denotes female ($f$) and male ($m$) speakers. These metrics are detailed next.

\subsection{Group Fairness Metrics}
\label{fairness}
The fairness metrics used in this study are derived from classification outcomes such as True Positives (TP), True Negatives  (TN), False Positives (FP), and False Negatives (FN), per gender group. Below, we provide a detailed statistical explanation for each fairness metric.\\

\noindent \textbf{Statistical Parity/Demographic Parity (SP)}: This measure evaluates the model ability to perform positive predictions (i.e., detection) within different demographic groups. For audio deepfake classification, if detection rates are equal for both genders, i.e., male and female, then the model's predictions exhibit demographic parity. While this measure can reveal the model's ability to detect audio deepfake within each group, it has the limitation of not reflecting the proportion of positive samples within each group, which can limit its interpretation of fairness. For SP, the proportion of positive predictions for a given group can be described as follows:

\begin{figure}[t]
    \centering
    \includegraphics[width=\linewidth]{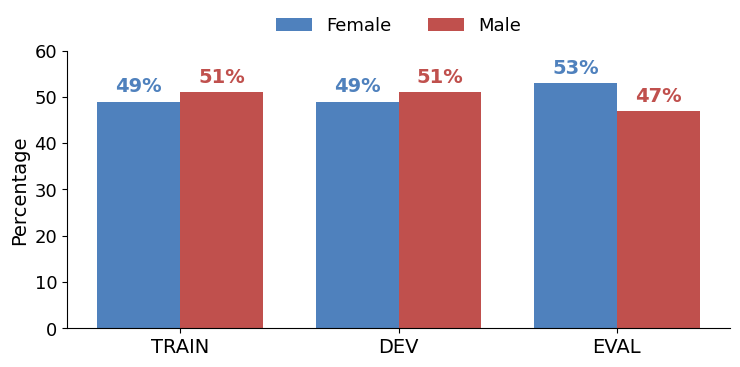}
    \caption{Balanced gender distribution for the ASVspoof 5 dataset across train, development, and evaluation splits.}
    \label{fig:gender_split}
\end{figure}

\begin{equation}
P(\hat{Y}\mid G ) = \frac{Count(\hat{Y} = 1 \mid G = g)}{Count(G=g)}  \quad \forall \, g \in \{f, m\}
\end{equation}

Note that parity is attained when the detection rate is the same for both groups, as described below:

\begin{equation}
P(\hat{Y} = 1 \mid g = f) = P(\hat{Y} = 1 \mid g = m)
\end{equation}

\begin{figure*}[t]
   \centering
   \includegraphics[width=0.9\linewidth]{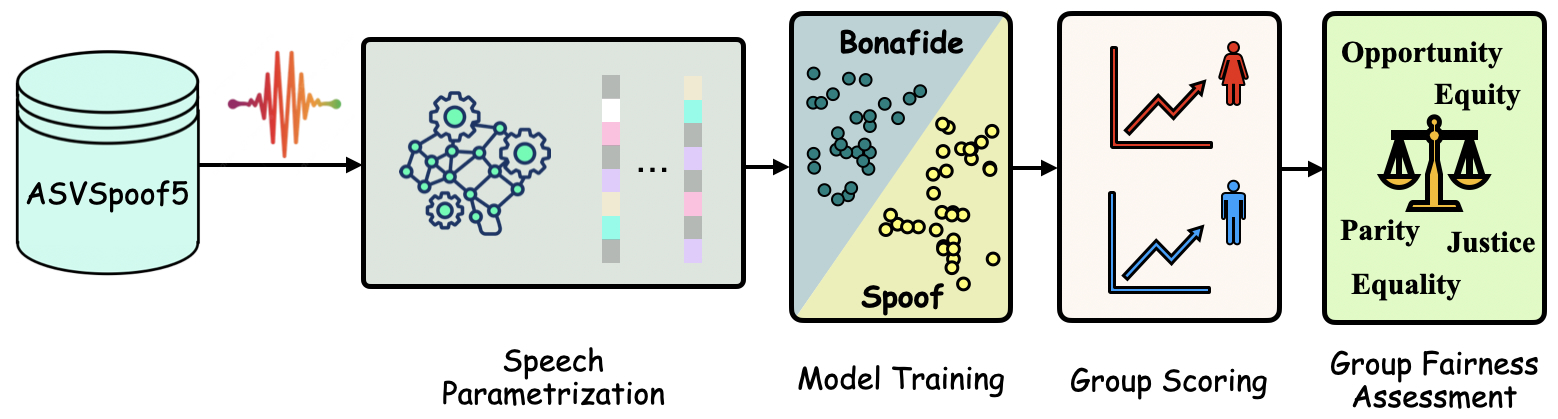}
   \caption{Illustration of the procedure adopted to perform group fairness assessment. Model performance is based on 5 group fairness measures.}
   \label{fig:exp_setup}
\end{figure*}

\noindent \textbf{Equal Opportunity (EOP)}: It evaluates the model ability to detect audio deepfake (i.e., positive predictions) within each group, but considering the samples with positive labels. Thus, EOP compares the detection rates for positive samples only in both male and female groups. If the detection rates for positive samples in both groups are equal, the model exhibits equality of opportunity. In such case, the model ensures that positive samples have the same likelihood of receiving a positive outcome regardless of demographic membership. For EOP, the proportion of positive predictions for a given group can be described as follows:

\begin{equation}
P(\hat{Y}\mid G ) = \frac{Count(\hat{Y} = 1 \mid Y = 1, G = g)}{ Count(Y = 1|G=g)}
\end{equation}

Thus, equal opportunity is achieved when the detection rate is the same for both groups while considering positive samples, as described below:

\begin{equation}
\begin{split}
P(\hat{Y} = 1 \mid Y = 1, g = f) = P(\hat{Y} = 1 \mid Y = 1, g = m) 
\end{split}
\end{equation}

\noindent \textbf{Equality of Odds (EO)}: While EOP focuses on evaluating the model performance on a specific class, equality of odds is a fairness measure that aims to assess whether a model predicts outcomes equally well for both the positive and negative classes. Unlike of EOP, both the true positive and false negative rates are expected to be the same for each group. Thus, it extends EOP by additionally requiring the false positive rate to be equal across groups, measuring fairness for both positive and negative outcomes simultaneously.

\begin{equation}
P(\hat{Y}\mid G ) = \frac{Count(\hat{Y} = \hat{y} \mid Y = y, G = g)}{ Count(Y = y \mid G=g)}  \quad \forall \, y \in \{+, -\}
\end{equation}

\begin{equation}
\mathrm{EO}_g = P(\hat{Y}=1 \mid Y=0, G=g)
= \frac{FP_g}{FP_g + TN_g}
\end{equation}

\noindent \textbf{Predictive Parity (PP)}: It evaluates the precision of positive predictions for a gender group, revealing whether the precision (Positive Predictive Value (PPV)) is equal across groups, ensuring that a positive prediction has the same reliability for different demographic groups. This metric may conflict with the equality of odds when base rates differ across groups. 

\begin{equation}
\mathrm{PPV}_g = P(Y=1 \mid \hat{Y}=1, G=g)
\end{equation}

\noindent \textbf{Treatment Equality (TE)}: It measures the balance of false positives and false negatives by comparing their ratios across demographic groups. It emphasizes balance in error types rather than absolute error rates, making it particularly relevant when different misclassification errors incur different costs. Here, a high TE indicates that the model produces more false acceptances than false rejections for a given gender group.

\begin{equation}
\mathrm{TE}_g = \frac{FP_g}{FN_g}
\end{equation}

Note that this metric is useful when FP and FN have different real-world costs (e.g., false rejection of bonafide users vs. missed spoof attacks).

\section{Experimental Setup}
\subsection{Audio Deepfake Dataset}
All experiments are conducted using the ASVspoof 5 dataset \cite{c108}, which is the fifth edition in a series of ASVSpoof challenges. Compared to earlier editions, such as ASVspoof 2019 and ASVspoof 2021, ASVspoof 5 provides a near-balanced distribution of male and female speakers, as shown in Fig.~\ref{fig:gender_split}, enabling a detailed analysis of gender-related performance disparities in audio deepfake detection. The dataset contains both bonafide and AI-generated speech based on state-of-the-art text-to-speech (TTS) and voice conversion (VC) techniques \cite{c108}. In our experiments, we follow the official ASVspoof 5 evaluation protocol and use the predefined training (train), development (dev), and evaluation (eval) splits. Gender labels and class annotations are obtained directly from the protocol files. 

\subsection{Feature Extraction}
We investigate four feature representations, including two traditional handcrafted features and two self-supervised embeddings. All audio signals are resampled to 16 kHz and standardized to a fixed duration of 4.0 seconds. Signals shorter than 4 seconds are zero-padded, while longer signals are truncated. These pre-processing steps ensure a consistent temporal context across all feature representations and enables fair comparison between different feature extraction methods. 

The conventional acoustic features include the Log-Spectrogram (LogSpec) and the Constant-Q Transform (CQT). LogSpec provides a time-frequency representation of speech by applying a logarithmic transformation to the magnitude spectrogram, capturing energy variations across frequency bands \cite{c109}. CQT uses logarithmically spaced frequency bins, which emphasize pitch and harmonic structures that are particularly informative for detecting synthetic and manipulated speech \cite{c110}. The deep learning-based feature representations include WavLM and Wav2Vec~2.0 embeddings. WavLM is a self-supervised speech representation model trained on large-scale speech corpora, producing contextualized frame-level embeddings that encode acoustic and linguistic characteristics \cite{c111}. Wav2Vec~2.0 is another widely used self-supervised model that learns contextual speech representations directly from raw audio and has demonstrated strong performance in speech-related tasks \cite{c112}. All features are pre-extracted and stored as fixed-size two-dimensional time-frequency tensors, ensuring that the classifier operates on uniform-length feature representations, with no additional temporal segmentation applied during training or evaluation.

\subsection{Classifier and Training Protocol}
A ResNet-18 architecture is adopted as baseline classifier. To ensure a fair comparison across all features, the same model architecture and training configuration are used throughout our experiments. The AASIST model is also included, serving as a more recent reference model. It was trained using its original architecture, as described in \cite{AASIST2025}. As shown in Fig.~\ref{fig:exp_setup}, we follow a unified pipeline comprising feature extraction, classifier training, and gender-wise evaluation using fairness metrics. The model optimization is based on the AdamW optimizer, with the learning rate set to $3\times10^{-5}$ and a weight decay of $1\times10^{-4}$. To address the class imbalance, class-weighted cross-entropy loss is applied, where each class weight is computed. Training is performed using solely the ASVspoof 5 training split, and early stopping with patience of 15 is applied based on the validation loss computed on the development split to avoid overfitting. The best model checkpoint is selected based on minimum development loss, and a ReduceLROnPlateau scheduler with a reduction factor of 0.5 and patience of 5 epochs is used to adapt the learning rate during training.

\subsection{Evaluation Protocol}
For evaluation, the model checkpoint corresponding to the lowest validation loss is selected and applied to the ASVspoof 5 evaluation set. To analyze gender-related performance differences, the evaluation data is partitioned into three subsets: Female-only, Male-only, and a Combined set containing both genders (All). Class labels and gender information are obtained from the official evaluation protocol.

% \begin{table}
% \centering
% \caption{Performance in terms of EER (\%) and AUC (\%) of four speech representation based combined with a ResNet classifier.}
% \label{tab:eer_auc}
% \begin{tabular}{l l c c}
% \hline
% \textbf{Method} & \textbf{Group Evaluated} & \textbf{EER (\%)} & \textbf{AUC (\%)} \\
% \hline
% CQT     & ALL    & 45.0 & 56.8 \\
% CQT     & Female & 45.6 & 56.5 \\
% CQT     & Male   & 45.0 & 56.4 \\
% \hline
% LogSpec   & ALL   & 39.70 & 63.85 \\
% LogSpec  & Female & 39.54 & 64.10 \\
% LogSpec   & Male   & 39.79 & 63.62 \\
% \hline
% Wav2vec & ALL    & 21.5 & 85.9 \\
% Wav2vec & Female & 21.8 & 85.7 \\
% Wav2vec & Male   & 21.2 & 86.1 \\
% \hline
% WavLM   & ALL    & 48.8 & 52.0 \\
% WavLM   & Female & 47.9 & 52.7 \\
% WavLM   & Male   & 49.6 & 50.5 \\
% \hline

% \end{tabular}
% \end{table}

\begin{table}
\caption{Fairness assessment based on the statistical parity metric.}
\label{tab:selection_rate}
\centering
\begin{tabular}{l|r|r|r|l}
\hline
\textbf{Model} & \textbf{Female} & \textbf{Male} & \textbf{Diff (F-M)} & \textbf{p-value (Holm)} \\
\hline
AASIST      & 0.190  & 0.210 & -0.016    & $<1\times10^{-16}$ \\
CQT      & 0.380  & 0.290 & 0.090    & $<1\times10^{-16}$ \\
LogSpec        & 0.324 & 0.334 & -0.009 & $<1\times10^{-16}$  \\
Wav2vec  & 0.349  & 0.316 & 0.033    & $<1\times10^{-16}$ \\
WavLM    & 0.195  & 0.182 & 0.011    & $<1\times10^{-16}$ \\
\hline
\end{tabular}

\end{table}

\begin{table}
\caption{Fairness assessment based on the equal opportunity metric.}
\label{tab:equal_opportunity}
\centering
\begin{tabular}{l|r|r|r|l}
\hline
\textbf{Model} & \textbf{Female} & \textbf{Male} & \textbf{Diff (F-M)} & \textbf{p-value (Holm)} \\
\hline
AASIST     & 0.172  & 0.197 & -0.007    & $<1\times10^{-16}$ \\
CQT      & 0.474  & 0.360 & 0.114    & $<1\times10^{-16}$ \\
LogSpec  & 0.498 & 0.501 & -0.003 & 0.2171 \\
Wav2vec  & 0.660  & 0.631 & 0.029    & $<1\times10^{-16}$ \\
WavLM    & 0.543  & 0.511 & 0.032    & $<1\times10^{-16}$ \\
\hline
\end{tabular}
\end{table}

During inference, the model outputs logits for the spoof and bonafide classes, which are converted into posterior probabilities using the Softmax function. The posterior probability of the bonafide class is used as the detection score. An operating threshold is derived on the development set at the EER operating point and then applied uniformly to the Female, Male, and All (combined) evaluation subsets, ensuring consistent operating conditions across feature representations. This development-set threshold is used to compute all fairness metrics, so that gender disparities are assessed at the same operating point. In this work, EER is used only to define a common operating threshold, while the analysis focuses on group-wise fairness metrics computed at that threshold. Statistical significance is assessed using z-tests with Holm correction across all comparisons ($\alpha=0.05$).

\subsection{Statistical Analysis}
\label{stat_analysis}
To assess gender-related disparities in audio deepfake detection, we conducted a statistical significance analysis across fairness metrics, including Statistical Parity, Equal Opportunity, Equality of Odds, Predictive Parity, and Treatment Equality. For each model and metric, we computed the difference between female and male groups:

\begin{equation}
\Delta_{f-m} = \mathrm{Metric}_{f} - \mathrm{Metric}_{m}.
\label{eq:delta}
\end{equation}

Two-proportion z-tests are performed for each metric, treating the metric estimates as proportions. The resulting z-statistics were used to compute p-values, which were then corrected for multiple comparisons using the Holm-Bonferroni procedure to control the gender-wise error rate.

Only metrics with p-values below the significance threshold of 0.05 after Holm correction were considered statistically significant. For each model and fairness metric, the null hypothesis ($H_0$) states that the metric value is equal for the female and male groups, indicating that the observed disparities are unlikely to have arisen by chance. %This rigorous statistical testing ensures that our fairness analysis reflects true model behavior rather than random variation in the dataset.

\section{Results and Discussion}

In this section, we present our analysis regarding gender-based fairness in audio deepfake detection. 

\begin{table}
\caption{Fairness assessment based on the equality of odds metric.}
\label{tab:equality_of_odds}
\centering
\begin{tabular}{l|r|r|r|l}
\hline
\textbf{Model} & \textbf{Female} & \textbf{Male} & \textbf{Diff (F-M)} & \textbf{p-value (Holm)} \\
\hline
AASIST      & 0.279  & 0.358 & -0.078    & $<1\times10^{-16}$ \\
CQT      & 0.357  & 0.271 & 0.086    & $<1\times10^{-16}$ \\
LogSpec & 0.282 & 0.289 & -0.007 & $<1\times10^{-16}$  \\
Wav2vec  & 0.273  & 0.231 & 0.042    & $<1\times10^{-16}$ \\
WavLM    & 0.110  & 0.096 & 0.014    & $<1\times10^{-16}$ \\
\hline
\end{tabular}
\end{table}

\begin{table}
\caption{Fairness assessment based on the predictive parity metric.}
\label{tab:predictive_parity}
\centering
\begin{tabular}{l|r|r|r|l}
\hline
\textbf{Model} & \textbf{Female} & \textbf{Male} & \textbf{Diff (F-M)} & \textbf{p-value (Holm)} \\
\hline
AASIST      & 0.679  & 0.687 & -0.007     & 0.003 \\
CQT      & 0.244  & 0.263 & -0.019     &$<1\times10^{-16}$ \\
LogSpec  & 0.301 & 0.318 &  -0.017 & $<1\times10^{-16}$ \\
Wav2vec  & 0.371  & 0.424 & -0.053    & $<1\times10^{-16}$ \\
WavLM    & 0.545  & 0.5891 & -0.044     & $<1\times10^{-16}$ \\
\hline
\end{tabular}
\end{table}
\subsection{Analysis I - Statistical Parity}

Table~\ref{tab:selection_rate} presents the results for Statistical Parity (SP), which measures whether the overall rate of positive predictions, i.e., spoof detections, is comparable across genders, regardless of the true label distribution. Large SP gaps indicate that one group is flagged as spoof more often than the other, which can translate into uneven user experience in real deployments (e.g., one group being challenged more frequently). Results show that the direction and magnitude of SP disparity varies by feature and model. CQT, for instance, exhibits the largest gap, with a higher positive prediction rate for female speech, i.e., 0.380 versus 0.290 for male (+0.090). Wav2Vec and WavLM also predict spoof slightly more often for females, achieving, respectively, +0.033 and +0.011 compared to males. Differently, AASIST and LogSpec show slightly higher spoof prediction rates for males, providing, respectively, -0.016 and -0.009 for male. Although several differences are numerically small, p-values indicate that these gaps are statistically significant for all tested systems.

\subsection{Analysis II - Equal Opportunity}

Table~\ref{tab:equal_opportunity} summarizes the results for Equal Opportunity (EOP), which focuses on the true positive rate (TPR) for spoof samples, as described in \ref{method}. The goal is to assess whether AI-generated speech has the same chance of being detected for female and male speakers, with lower EOP for a group implying that attacks targeting that group are more likely to bypass the detector. Results for CQT represent the strongest disparity in TPR, favoring female speech (i.e., 0.474) over male speech (i.e., 0.360), with a difference of +0.114. Wav2Vec and WavLM also yield higher TPR for females, providing, respectively, a difference of +0.029 and +0.032 over males. This suggests that spoof artifacts may be more detectable for female voices under these representations or that the model is more sensitive to female-specific cues. For the AASIST, a small advantage was found for males with a -0.007 difference. Importantly, LogSpec is the only configuration where the EOP gap is not statistically significant, with p=0.2171, indicating comparable spoof-detection opportunity across genders under this feature representation.

\begin{table}
\caption{Fairness assessment based on the treatment equality metric.}
\label{tab:treatment_equality}
\centering
\begin{tabular}{l|r|r|r|l}
\hline
\textbf{Model} & \textbf{Female} & \textbf{Male} & \textbf{Diff (F-M)} & \textbf{p-value (Holm)} \\
\hline
AASIST  & 0.0977  & 0.0991 & -0.0014    & $<1\times10^{-16}$ \\
CQT      & 2.7904  & 1.574 & 1.216    & $<1\times10^{-16}$ \\
LogSpec  & 0.734 & 0.749 & -0.015 & $<1.32\times10^{-16}$ \\
Wav2vec  & 3.311  & 2.331 & 0.979    & $<1\times10^{-16}$ \\
WavLM    & 0.993  & 0.729 & 0.263   & $<1\times10^{-16}$ \\
\hline
\end{tabular}
\end{table}

\subsection{Analysis III - Equality of Odds (EO)}

Results for Equality of Odds are presented in Table~\ref{tab:equality_of_odds}. The measure requires parity in both TPR and FPR across groups. False positive rate (FPR), as defined in Eq. 6, captures how often bonafide audio is incorrectly flagged as spoof. This is a key fairness dimension in biometric settings because it reflects unequal friction, with higher FPR for a group meaning that more legitimate users from that group are incorrectly rejected. Results show heterogeneous behavior. AASIST produces a substantially higher FPR for male speakers (i.e., 0.358 versus 0.279), indicating that bonafide male speech is more frequently misclassified as spoof under this model. In contrast, CQT yields higher FPR for females, with a  +0.086 difference over males. LogSpec produces nearly matched FPR across groups (i.e., -0.007), while WavLM achieves the lowest FPR overall, with 0.110 for female and 0.096 for male, with a small gap of +0.014. All gaps in Table~\ref{tab:equality_of_odds} are statistically significant, suggesting that even modest FPR differences are systematic.

\subsection{Analysis IV -  Predictive Parity}

Table~\ref{tab:predictive_parity} provides the results for Predictive Parity (PP), which measures whether positive predictions are equally reliable across genders, i.e., whether “spoof” predictions have similar precision for female and male speech. Unlike EOP (which reflects missed attacks), PP captures the likelihood that a flagged sample is truly spoof, affecting the perceived trustworthiness of decisions across groups. Results in Table~\ref{tab:predictive_parity} show higher PP for male speakers for every model, with all differences being negative, which implies that, across all evaluated systems, a “spoof” prediction is more likely to be correct for male speech than for female speech. The magnitude varies: the smallest gap is observed for AASIST (i.e., -0.007), while the largest appears for Wav2Vec (i.e., -0.053), followed by WavLM (i.e., -0.044). This cross-model consistency indicates that the effect is not specific to one architecture or feature, but may reflect broader score-distribution differences between genders at the chosen operating point.

\subsection{Analysis V - Treatment Equality}

Table~\ref{tab:treatment_equality} presents the results for Treatment Equality (TE), which evaluates the balance between error types by comparing the ratio FP/FN within each group, with values above 1 indicating relatively more false positives than false negatives and values below 1 indicate the opposite. Our results highlight strong model-dependent differences. CQT and Wav2Vec exhibit very high TE values, especially for females, achieving, respectively, 2.79 and 3.31. This indicates that these systems generate substantially more false positives than false negatives for female speakers. By contrast, AASIST has TE near 0.1 for both groups, indicating that its errors are dominated by false negatives (missed spoofs) rather than false positives, although with minimal gender difference of -0.001. WavLM is closer to balance, providing 0.993 for female and 0.729 for male, with +0.263 difference, while LogSpec remains relatively stable across genders, with -0.015 difference. All TE gaps are statistically significant, confirming that error-type asymmetry differs by gender.

\section{Conclusion}
This paper presented a gender-focused fairness evaluation of audio deepfake detection on the ASVspoof 5 benchmark. We compared a ResNet-18 detector trained with four feature representations (LogSpec, CQT, Wav2Vec 2.0, and WavLM) against the AASIST reference model, and assessed model behavior using five group fairness metrics computed at a common operating point (threshold set on the development set at the EER operating point). Our results reveal statistically significant gender disparities across multiple fairness criteria, showing that aggregate operating-point summaries such as EER do not explicitly characterize how errors are distributed across demographic groups. Across the evaluated systems, fairness outcomes are strongly model- and feature-dependent: CQT exhibits the largest gender gaps across several metrics, whereas AASIST and LogSpec show comparatively smaller disparities. Moreover, Predictive Parity consistently favors male speakers across all models, suggesting that the reliability of “spoof” decisions differs by gender under a fixed threshold. Overall, these findings reinforce the need to complement standard performance evaluation with fairness-aware analysis when developing and selecting audio deepfake countermeasures for deployment.

Future work should investigate the underlying causes of these disparities (e.g., score distribution shifts, feature sensitivity to pitch-related characteristics, or subgroup-specific spoof artifacts) and evaluate mitigation strategies.

%\section{LIMITATION AND FUTURE WORK}

% While our study provides a systematic evaluation of gender fairness across multiple audio deepfake detection features and models, it is limited to binary gender categories, gender-balanced datasets, and a fixed set of architectures, and does not isolate performance by specific attack types or explore fairness mitigation strategies. Building on these insights, future work will investigate the effects of imbalanced gender data, assess feature-specific vulnerability to different attack types, and explore advanced architectures, including self-supervised learning models, to improve both detection performance and demographic fairness, ultimately guiding the design of more equitable and robust audio deepfake detection systems.

% \section*{ACKNOWLEDGMENT}
% During the preparation of this work, the authors used Grammarly (Grammarly,
% Inc., San Francisco, CA, USA) in order to improve the readability and language quality of the manuscript. 


\begin{thebibliography}{99}

\bibitem{c1} Y. Bengio, I. Goodfellow, and A. Courville,{Deep Learning}, vol. 1. Cambridge, MA, USA: MIT Press, 2017.

\bibitem{c2}
J.~Yi, C.~Wang, J.~Tao, X.~Zhang, C.~Y.~Zhang, and Y.~Zhao,``Audio deepfake detection: A survey,''{arXiv preprint arXiv:2308.14970}, 2023.

\bibitem{c3}
L. Verdoliva,
``Media forensics and deepfakes: An overview,''
\emph{IEEE Journal of Selected Topics in Signal Processing},
vol. 14, no. 5, pp. 910--932, Aug. 2020.

\bibitem{c4}
Z. Wu, J. Yamagishi, T. Kinnunen, C. Hanilci, M. Sahidullah, A. Sizov, N. Evans, M. Todisco, and H. Delgado,
``ASVspoof: The automatic speaker verification spoofing and countermeasures challenge,''
\emph{IEEE Journal of Selected Topics in Signal Processing},
vol. 11, no. 4, pp. 588--604, Jun. 2017.

\bibitem{c7}
S. Barocas, M. Hardt, and A. Narayanan,
\emph{Fairness and Machine Learning: Limitations and Opportunities}.
Cambridge, MA, USA: MIT Press, 2023.


\bibitem{c8}
C. Harris, C. Mgbahurike, N. Kumar, and D. Yang,
``Modeling gender and dialect bias in automatic speech recognition,''
in \emph{Findings of the Association for Computational Linguistics: EMNLP 2024},
pp. 15166--15184, 2024.


\bibitem{c9}
N. Mehrabi, F. Morstatter, N. Saxena, K. Lerman, and A. Galstyan,
``A survey on bias and fairness in machine learning,''
\emph{ACM Computing Surveys},
vol. 54, no. 6, pp. 1--35, 2021.


\bibitem{c10}
J. Buolamwini and T. Gebru,
``Gender shades: Intersectional accuracy disparities in commercial gender classification,''
in \emph{Proc. Conf. Fairness, Accountability, and Transparency},
pp. 77--91, 2018.

\bibitem{c11}
I. D. Raji, A. Smart, R. N. White, M. Mitchell, T. Gebru, B. Hutchinson, J. Smith-Loud, D. Theron, and P. Barnes,
``Closing the AI accountability gap: Defining an end-to-end framework for internal algorithmic auditing,''
in \emph{Proc. Conf. Fairness, Accountability, and Transparency},
pp. 33--44, 2020.

\bibitem{c12}
B. Zhang, H. Cui, V. Nguyen, and M. Whitty,
``Audio deepfake detection: What has been achieved and what lies ahead,''
\emph{Sensors},
vol. 25, no. 7, Art. no. 1989, 2025.

\bibitem{c13}
M. Todisco, X. Wang, V. Vestman, M. Sahidullah, H. Delgado, A. Nautsch, J. Yamagishi, N. Evans, T. Kinnunen, and K. A. Lee,
``ASVspoof 2019: Future horizons in spoofed and fake audio detection,''
\emph{arXiv preprint} arXiv:1904.05441, 2019.

\bibitem{c14}
S. Verma and J. Rubin,
``Fairness definitions explained,''
in \emph{Proc. Int. Workshop on Software Fairness},
2018.

\bibitem{c16}
M. Hardt, E. Price, and N. Srebro,
``Equality of opportunity in supervised learning,''
in \emph{Advances in Neural Information Processing Systems},
vol. 29, 2016.

\bibitem{c102}
M. Pu, M. Y. Kuan, N. T. Lim, C. Y. Chong, and M. K. Lim,
``Fairness evaluation in deepfake detection models using metamorphic testing,''in \emph{Proc. 7th Int. Workshop on Metamorphic Testing (MET)},
2022.


\bibitem{c106}
T. Estella, A. Zahra, and W.-K. Fung,
``Accessing gender bias in speech processing using machine learning and deep learning with gender balanced audio deepfake dataset,''
in \emph{Proc. 9th Int. Conf. Informatics and Computing (ICIC)},
2024.



\bibitem{c108}
X.~Wang, H.~Delgado, H.~Tak, J.~W. Jung, H.~J. Shim, M.~Todisco, J.~Yamagishi, \emph{et~al.}, “ASVspoof 5: Crowdsourced speech data, deepfakes, and adversarial attacks at scale,” \emph{arXiv preprint arXiv:2408.08739}, 2024.


\bibitem{c109}
S.~Davis and P.~Mermelstein,
``Comparison of parametric representations for monosyllabic word recognition in continuously spoken sentences,''
\emph{IEEE Trans. Acoust., Speech, Signal Process.},
vol.~28, no.~4, pp.~357--366, Aug.~1980.

\bibitem{c110}
J.~C.~Brown,
``Calculation of a constant Q spectral transform,''
\emph{J. Acoust. Soc. Am.},
vol.~89, no.~1, pp.~425--434, Jan.~1991.

\bibitem{c111}
S.~Chen, C.~Wang, Z.~Chen, Y.~Wu, S.~Liu, Z.~Chen, \emph{et~al.},
``WavLM: Large-scale self-supervised pre-training for full stack speech processing,''
\emph{IEEE J. Sel. Topics Signal Process.},
vol.~16, no.~6, pp.~1505--1518, Oct.~2022.

\bibitem{c112}
A.~Baevski, Y.~Zhou, A.~Mohamed, and M.~Auli,
``wav2vec~2.0: A framework for self-supervised learning of speech representations,''
in \emph{Adv. Neural Inf. Process. Syst. (NeurIPS)},
vol.~33, pp.~12449--12460, 2020.

\bibitem{c113}
K.~He, X.~Zhang, S.~Ren, and J.~Sun,
``Deep residual learning for image recognition,''
in \emph{Proc. IEEE Conf. Comput. Vis. Pattern Recognit. (CVPR)},
pp.~770--778, 2016.

\bibitem{c115}
V.~Nallaguntla, A.~Fursule, S.~Kshirsagar, and A.~R.~Avila,
``PhonemeDF: A Synthetic Speech Dataset for Audio Deepfake Detection and Naturalness Evaluation,''
in \emph{Proceedings of the 15th International Conference on Language Resources and Evaluation (LREC 2026)},
2026, submitted.



\bibitem{c116}
D.~E.~Temmar, A.~Hamadene, V.~Nallaguntla, A.~R.~Fursule,
M.~S.~Allili, S.~Kshirsagar, and A.~R.~Avila,
``Phonetic analysis of real and synthetic speech using HuBERT embeddings: Perspectives for deepfake detection,''
in \emph{Proc. IEEE Int. Conf. Systems, Man, and Cybernetics (SMC)},
2025, pp.~86--91, doi:~10.1109/SMC58881.2025.11343334.


\bibitem{c117}
Y.~Ju, S.~Hu, S.~Jia, G.~H.~Chen, and S.~Lyu,
``Improving fairness in deepfake detection,''
in \emph{Proc. IEEE/CVF Winter Conf. Appl. Comput. Vis. (WACV)},
pp.~4655--4665, 2024.


\bibitem{c118}
L.~Trinh and Y.~Liu,
``An examination of fairness of AI models for deepfake detection,''
\emph{arXiv preprint arXiv:2105.00558}, 2021.



\bibitem{c119}
N.~Alshareef, X.~Yuan, K.~Roy, and M.~Atay,
``A study of gender bias in face presentation attack and its mitigation,''
\emph{Future Internet},
vol.~13, no.~9, Art.~no.~234, Sep.~2021.


\bibitem{c120}
Z.~Wu, T.~Kinnunen, N.~Evans, J.~Yamagishi, C.~Hanil{\c{c}}i, M.~Sahidullah, and A.~Sizov,
``ASVspoof 2015: The first automatic speaker verification spoofing and countermeasures challenge,''
in \emph{Proc. Interspeech},
2015, pp.~2037--2041.

\bibitem{c121}
T.~Kinnunen, M.~Sahidullah, H.~Delgado, M.~Todisco, N.~Evans, J.~Yamagishi, and K.~A.~Lee,
``The ASVspoof 2017 challenge: Assessing the limits of replay spoofing attack detection,''
in \emph{Proc. Interspeech},
2017.


\bibitem{c122}
J.~Yamagishi, X.~Wang, M.~Todisco, M.~Sahidullah, J.~Patino, A.~Nautsch, and H.~Delgado,
``ASVspoof 2021: Accelerating progress in spoofed and deepfake speech detection,''
\emph{arXiv preprint arXiv:2109.00537},
2021.


\bibitem{c124}
N.~Mehrabi, F.~Morstatter, N.~Saxena, K.~Lerman, and A.~Galstyan,
``A survey on bias and fairness in machine learning,''
\emph{ACM Computing Surveys},
vol.~54, no.~6, pp.~1--35, 2021.

\bibitem{c125}
A.~V.~Nadimpalli and A.~Rattani,
``GBDF: Gender balanced deepfake dataset towards fair deepfake detection,''
in \emph{Proc. Int. Conf. Pattern Recognition (ICPR)},
Cham, Switzerland: Springer Nature Switzerland, Aug.~2022,
pp.~320--337.

\bibitem{agarwal2024fairdeepfake}
A.~Agarwal and N.~Ratha,
``Deepfake: Classifiers, fairness, and demographically robust algorithm,''
in \emph{Proc. IEEE 18th Int. Conf. Automatic Face and Gesture Recognition (FG)},
May~2024, pp.~1--9.
\bibitem{yadav2024fairssd}
Yadav, A. K. S., Bhagtani, K., Salvi, D., Bestagini, P., and Delp, E. J. (2024).
FairSSD: Understanding Bias in Synthetic Speech Detectors.
In \textit{Proceedings of the IEEE/CVF Conference on Computer Vision and Pattern Recognition}, pages 4418--4428.
\bibitem{bird2023real}
Bird, J. J. and Lotfi, A. (2023).
Real-time Detection of AI-Generated Speech for DeepFake Voice Conversion.
\textit{arXiv preprint arXiv:2308.12734}.
\bibitem{yang2025}
Yang, T., Sun, C., Lyu, S., and Rose, P. (2025).
Forensic Deepfake Audio Detection using Segmental Speech Features.
\textit{arXiv preprint arXiv:2505.13847}.
\bibitem{AASIST2025}
J.-W. Jung, H.-S. Heo, H. Tak, H.-J. Shim, J. S. Chung, B.-J. Lee, H.-J. Yu, and N. Evans, 
``AASIST: Audio Anti-Spoofing Using Integrated Spectro-Temporal Graph Attention Networks,'' 
in \textit{Proc. IEEE International Conference on Acoustics, Speech and Signal Processing (ICASSP)}, 
2022, pp. 6367--6371.

\bibitem{panda2026bias}
A. Panda, T. Ghosh, T. Choudhary, and R. Naskar,
``Bias-Free? An Empirical Study on Ethnicity, Gender, and Age Fairness in Deepfake Detection,''
\textit{ACM Computing Surveys}, 2026.

\end{thebibliography}
\end{document}